\documentclass[conference,10pt]{IEEEtran}
\IEEEoverridecommandlockouts

\usepackage{textcomp}
\usepackage{gensymb}
\usepackage{multirow}
\usepackage[ruled,linesnumbered,noend]{algorithm2e}

\usepackage{cite}
\usepackage{subcaption}
\usepackage{amsmath,amssymb,amsfonts}
\usepackage[ruled,linesnumbered,noend]{algorithm2e} 
\usepackage{graphicx}
\usepackage{textcomp}

\usepackage{tabularx}
\usepackage{makecell}

\usepackage{mathtools}
\DeclarePairedDelimiter{\ceil}{\lceil}{\rceil}

\usepackage{url}
\def\BibTeX{{\rm B\kern-.05em{\sc i\kern-.025em b}\kern-.08em
    T\kern-.1667em\lower.7ex\hbox{E}\kern-.125emX}}

\begin{document}

\title{SHINE (Strategies for High-frequency INdoor Environments) with Efficient THz-AP Placement
\thanks{This project was funded by CMU Portugal Program: CMU/TMP/0013/2017- THz Communication for Beyond 5G Ultra-fast Networks.}
}

\author{\IEEEauthorblockN{Rohit Singh \textsuperscript{1}, Douglas Sicker \textsuperscript{1,2}}
\IEEEauthorblockA{ \textsuperscript{1}  \textit{Engineering \& Public Policy}, \textit{Carnegie Mellon University}, Pittsburgh, USA \\
			       \textsuperscript{2}  \textit{School of Computer Science}, \textit{Carnegie Mellon University}, Pittsburgh, USA \\
Email: rohits1@andrew.cmu.edu, sicker@cmu.edu}
}

\maketitle

\section*{ABSTRACT}
The increasing demand for ultra-high throughput in ultra-dense networks might take a toll on 5G capacity. Moreover, with Internet-of-Things (IoT) and the growing use-cases for indoor killer-applications, it will be necessary to look beyond 5G technologies. One such promising technology is to move higher in the frequencies, such as the THz ($300$ $GHz$-$10$ $THz$) spectrum. THz has a massive number of greenfield-contiguous channels ranging from $10$ $GHz$ to $200$ $GHz$ (best case), which was not available in the traditional radio frequency (RF) or millimeter wave (mmWave) bands. Although THz has immense potential to cater to such demands, it comes with numerous challenges revolving around hardware, link budget, mobility, blockages, scheduling, and deployment. Terahertz access points (THz-APs) are sensitive to deployment and can critically impact a system\textquotesingle s dynamics (i.e., coverage, throughput, and efficiency). In this paper, we present Strategies for High-frequency INdoor Environments or SHINE, which focuses on efficient AP deployment in the THz spectrum and draws motivation from approaches used indoor to improve lighting conditions. Due to THz\textquotesingle s limited coverage area of a few meters, the number of THz-APs required to satisfy a densely populated room will be higher compared to today's single router/box/AP model. This increased number of THz-APs will not only increase the operational costs, but also (in some cases) can make the system inefficient. Through SHINE, we explore the deployment-related challenges and propose strategies to mitigate the same. 

\section*{Keywords}  Terahertz (THz), Beyond 5G (B5G), Indoor AP Placement, Spectrum Efficiency, Small-scale mobility, Mobile Blockage.

\section{Introduction}

The current 5G network is envisioned to cater to the growing user demands for higher throughput and lower latency applications \cite{ITUR}. 5G is supposed to use a combination of mid ($<6GHz$) and high ($30-300GHz$) frequency bands, which can provide contiguous bandwidth of $425$ $MHz$ and $7$ $GHz$, respectively \cite{OurTPRC}\cite{B100GHz}. These bandwidths, coupled with state-of-the-art modulation schemes, coding schemes, and access networks, can provide the system with a few $Gbps$ of throughput \cite{5gReview}. However, in the future, killer-applications, such as virtual reality (VR), mixed-reality (MR), holographic-type communications (HTC), digital twin, tactile internet, and Internet of Nano Things (IoNT), might require a peak throughput of $1$ $Tbps$, average throughput of $>100$ $Gbps$, and latency $<0.1$ $ms$ \cite{TeraCom}. These demands will overwhelm the soon to be deployed 5G network. It is time that we look beyond 5G (B5G), specifically beyond mmWave bands, and move higher in the frequency. Although THz might not be deployed until 2030 or later, we should explore the application of this spectrum now.



Although there exist technologies that can achieve a throughput of a few $Gbps$ \cite{B100GHz}, it still cannot meet the performance that THz alone can achieved. However, compared to lower frequency bands, THz has a distance issue of limited cell coverage of $<10$ $m$ for direct Line-of-sight (LOS) communication \cite{OurTPRC}. This might result in a need for an ultra-small cell deployment strategy and increased capital expenditure (CAPEX). Nevertheless, the relatively broader coverage of the RF bands makes it harder for the APs to reuse frequency without causing interference and making it unsuitable for ultra-dense networks (UDNs). On the other hand, 6G is envisioned to cater to extremely dense networks (EDNs) \cite{UDNSurvey}, which suggests deployment of more than $1000$ devices per $100$ $m^2$ \cite{TeraCom}. THz\textquotesingle s inherent distance issue mitigates this interference related problem \cite{OurTPRC}, making it ideal for EDNs.

Even if we limit the application of THz to indoor use only, there still exist deployment-related challenges. Resource management is already difficult \cite{OurIdVTC} in current systems and will become more complicated in the future. THz demands significant infrastructure, which in turn requires reliable resource allocation strategies. For example, the THz links need to be highly reliable for VR/MR applications, where repeated delays can make a user nauseate \cite{VRNausea}, or for health and public safety-related applications, which can result in loss of life. Even small-scale mobility (i.e., change in device orientation due to human body movement) and blockages can cause significant outages \cite{SSMNano}\cite{OurCCNC}\cite{THzBlk}, resulting in reduced throughput, frequent handoffs among THz-APs, inadequate resource allocation, and reduced system efficiency. It will be vital that we add intelligence to these antennas \cite{OurCCNC} \cite{SerLig} for faster beam alignment. It is worth noting that beam alignment will become complicated with increasing number of devices. In this paper, we explore THz-APs placement strategies as a solution to the challenges mentioned above.


\subsection{Our Contribution }

In this paper, we present Strategies for High-frequency INdoor Environments or SHINE, which focuses on efficient deployment in the THz spectrum. Through SHINE, we show that by changing the THz-AP constellation and density in a room, we can improve system efficiency (spectrum and energy) both with and without blockages. SHINE proposes that for a given room dimension, there exists a particular type of THz-AP constellation and density that can optimize the system efficiently. To improve energy efficiency, we keep the total usable power by all the APs in a room under an acceptable threshold, which can be based on hardware limitations and safety standards \cite{OurTPRC}. We show that even with transmit power limitations, SHINE can efficiently design THz-APs deployment for a room. 

To deploy these THz-APs efficiently, we draw motivation from approaches used indoor to improve lighting conditions, where the intensity of light can be considered as the throughput. Generally light sources are placed on the ceilings (center of the room as a single source and distributed is a grid) or on the perimeter (wall fixtures, standing lights, and table lamps) of the room. Similar placement strategies have been proposed by researchers while conducting THz simulation, such as homogeneous ceiling distribution \cite{CovTHz}, and for the perimeter model THz Plug \cite{ThzPlug} and reconfigurable intelligent surfaces \cite{ReconSur}. However, to the best of our knowledge, none of the research has categorized and performed a detailed analysis for such placement types. SHINE shows that by changing the design and density of THz-APs, the systems can improve efficiency with minimal energy, which is a critical aspect of THz communication \cite{ESEffi}. Although in this paper we fix the THz-APs locations, which is the case for most light sources indoor, we can also extend SHINE for dynamic AP placement. Nevertheless, this dynamic placement will require reliable indoor wireless backhaul, which can become challenging. The SHINE model also discusses the idle room dimension for a range of THz frequencies and its corresponding AP placement settings. SHINE also considers small-scale user mobility and mobile blockages to decide on the AP constellation. To measure spectrum and energy efficiency, we use user coverage, throughput, and the AP idle time as the evaluation metric.  

The rest of the paper is organized as follows. In Section \ref{SArch}, we discuss the propose strategies for AP placement with respect to the room dimension. In Section \ref{SPara}, we discuss the system model and the parameters for the proposed placement strategies. In Section \ref{SEval}, we evaluate SHINE for multiple scenarios and summarize our findings in Section \ref{Con}. 


    
\section{SHINE Architecture} \label{SArch}

In this Section, we discuss strategies for efficient THz-APs constellation deployed within a room. Not every square foot of a room is the same in terms of needing coverage. Both the user demand and the blockage distribution will vary based on the location within a room. Moreover, user mobility can result in severe outage scenarios and an increased probability of encountering blockages. Thus, an efficient AP deployment strategy is required. Efficient coverage for dense networks have been extensively studied in wireless sensor network (WSN) \cite{SensMOP}. Although THz networks will have demands similar to WSN, i.e., limited coverage and a limited energy budget, it will have a significant difference that makes the traditional sensor deployment approaches inefficient. For example, most of the devices in WSN are mobile and can optimize the system by moving around, which is not the case in THz, where the APs are likely to be static. These THz-APs will require a backhaul with optical fiber to support the high speed fronthaul, which is challenging in distributed THz networks. The deployment strategies studied in WSN will not be applicable in THz systems. Besides, considering a random constellation deployment of THz-AP will not guarantee an efficient strategy for a dense user model.

\setlength{\intextsep}{0pt}
\setlength\belowcaptionskip{-0.2 in}
\begin{figure}[t]
\centering
\includegraphics[width=3.5 in,height=2.5 in]{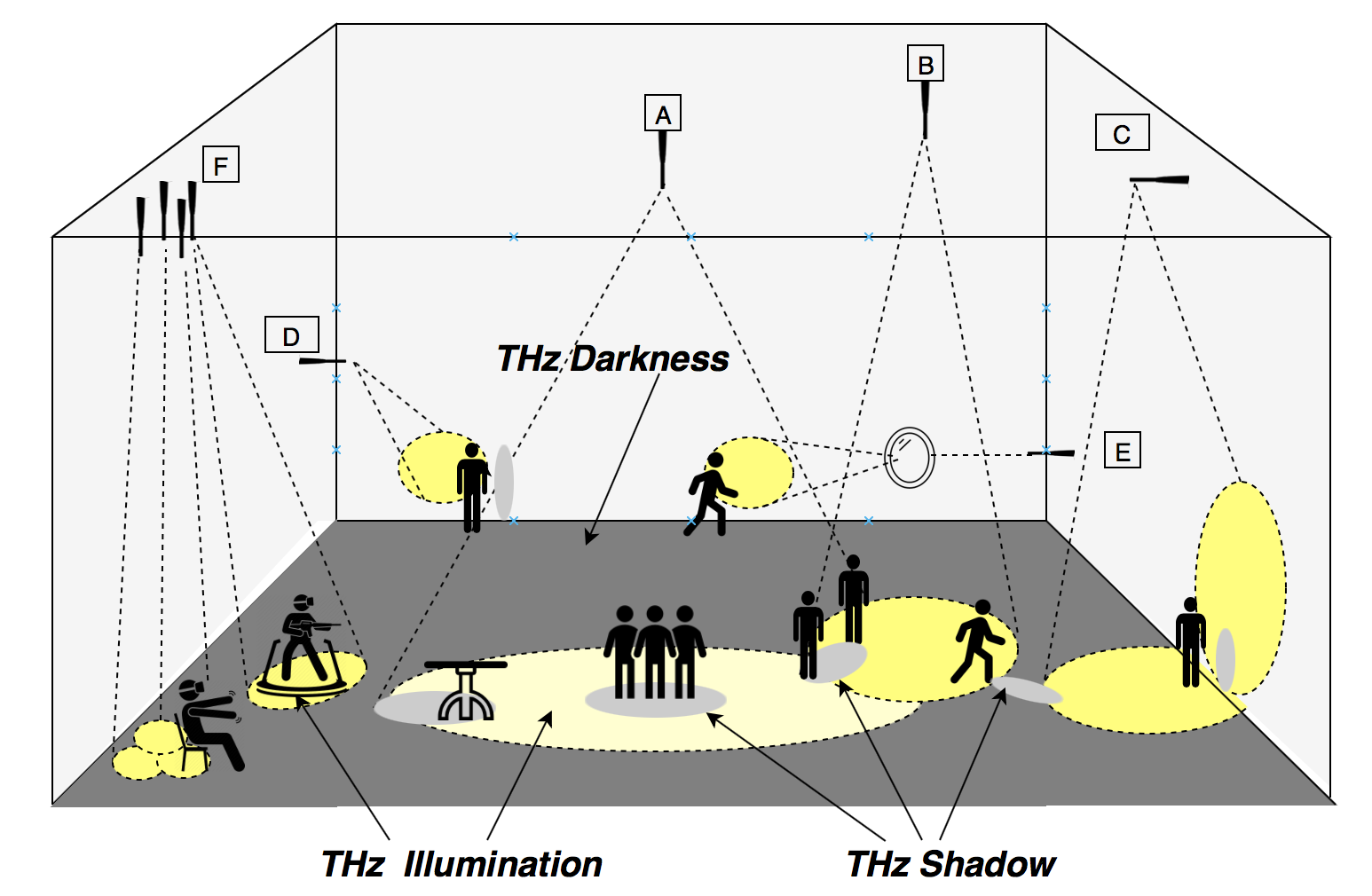}
\caption{THz-AP placement strategies.}
\label{IllSet}
\end{figure}

The THz-AP deployment problem is analogous to ``efficiently illuminating a room with light bulbs."  For example, allocating too few bulbs will not help illuminate the room effectively and will result in poor lighting, while too many light bulbs can lead to \textit{light pollution} (unnecessary or too much light), increased costs, and inefficiency. Based on this analogy, we try to design an efficient indoor deployment model for the THz spectrum.  To understand the demands for different areas of a THz enabled room we classify a room into 3 different regions:

\begin{itemize}
\item THz Darkness: In this region of the room, the signal strengths originated from the THz-APs are too weak to provide the bare minimum requested data rate, as represented by black in Fig. \ref{IllSet}. Due to short-range coverage, variations in environment, such as high temperature and high humidity, and limited transmit power, a room can have multiple dark regions.

\item THz Illumination: The rest of the region in a room, which doesn\textquotesingle t fall in the THz-Darkness regions, can be considered as the Illumination region, as represented by yellow in Fig. \ref{IllSet}. In these regions, the THz signal is strong enough to satisfy more than the minimum requested data rate. The intensity of the yellow patches can vary based on the AP placement strategy discussed later.  

\item THz Shadow: This region of the room is similar to the THz-Darkness regions and also cannot provide the bare minimum requested data rate. However, these regions are located in the THz-Illumination region, which can result in reduced average illumination or throughput. The THz-Shadow regions originate due to blockages (static or mobile) and mobility induced outages, as represented by grey in Fig. \ref{IllSet}. Even if a user is very close to a THz-AP, due to the low penetration power of THz, the user can experience a temporary shadow due to the presence of furniture and even other user(s) standing nearby. For dense mobile users, these grey patches vary in size and location, and can be mitigated by tracking user mobility.

\end{itemize}


SHINE proposes six different forms of AP Placement, as shown in Fig. \ref{IllSet}. THz-APs can be on the ceiling either at the center, i.e., Type A, like traditional lights, or spread across the room in a grid, i.e., Type B, like LED lights used indoors. Type B placement is more focused and intense compared to Type A placement as represented by the intensity of yellow in Fig. \ref{IllSet}.  Moreover, THz-APs can also be placed on the perimeter of the room or side of the walls, like tube-light or wall lights, i.e., Type C, or close to a user (like a hanging light or stand-alone light) , i.e., Type D. A variation of the Type D model will be Type E, where the APs are near or lower than the usual human height, like a table lamp. Type E is a more dedicated form of communication and might require non-line-of-sight (NLOS) communication to avoid outages. The efficiency of this type of placement is dependent on the material and the operating frequency \cite{THzMat}. Similar dedicated communication can also be done through the ceiling deployment, i.e., Type F, which is multiple closely deployed APs. The number of APs required for each placement type is subjected to the user distribution, likelihood of user and blockage location, and efficiency of the placement type, which is evaluated later in the paper. 




\section {SHINE Parametric  Model} \label{SPara}

In this Section, we define the system parameters for the THz-AP architecture proposed in Section \ref{SArch}.

\subsection{System Model} \label{SysModel}

In this Section, we briefly discuss the system models and assumptions that are applied in this paper. Table \ref{Tab} summarizes a list of notations and simulation values used in the paper.

To evaluate the effectiveness of the AP placement types shown in Fig. \ref{IllSet}, we need to understand the region of THz-Illumination or the available data rate for a single THz-AP. The achievable data rate at the $j^{th}$ user from the $i^{th}$ THz-AP is shown in Equation \ref{RateEq}, with the parameters explained in Table \ref{Tab}. We assume an antenna with conical shaped main lobe and uniformly illuminated circular aperture with $G(\delta)=\frac{52525}{\delta^2}$ \cite{OurCCNC}. The antenna beams are allowed to realign themselves when the throughput is $<\mathcal{R}^*$, which takes $t_{align}$ time. We assume that for perfect alignment the transmitter and receiver will have the same gain $G_t(\delta)=G_r(\delta)$.

\abovedisplayskip=-4pt
\belowdisplayskip=4pt
\begin{eqnarray}
R_{ij}=B_i log(1+\frac{P_t*G_i (\delta)*G_j (\delta)}{\mathcal{L}_T (f_c,d_{ij},\rho,T)* N_f*B_i})
\label{RateEq}
\end{eqnarray}

THz obeys most optical properties, such as it can reflect, refract, scatter, and can be blocked. Selective materials, such as paper, Teflon, and plastic, might allow THz signal to propagate with some attenuation \cite{THzMat}. However, in the case of deployment, it is unlikely that the surfaces will be THz-friendly. Therefore, we consider that the THz signal will be blocked by all objects encounter in the link. The total path loss encountered by a THz link is $\mathcal{L}_T=(4 \pi d_{ij} f_c)/c)^2 * e^{\tau(f_c,\rho,T)d_{ij}} $, where the notations are explained in Table \ref{Tab}.

For a dense deployment of THz-APs, the total power consumption in a room will also grow. The THz network will be sensitive to energy use and will need methods for efficient use of power \cite{ESEffi}. We assume that the total power consumed by a room is fixed to $\mathcal{P}_o$ and is equally distributed between the number of THz-APs as $\mathcal{P}_t$. Although the denser deployment of APs will be better for THz-Illumination, this power constraint will prevent light pollution. Since the transmit power keeps on decreasing with the increase in the number of THz-APs, the system will converge for the placement strategies depending on the placement types and environment settings.

\setlength{\intextsep}{0pt}
\setlength\belowcaptionskip{0 in}
\setlength\abovecaptionskip{0 in}
\begin{table}[t]
\centering
\caption{Notations and Simulation Values used.}
\begin{tabular}{ |c|c|c| } 
 \hline
 Parameter & Description & Value \\ \hline \hline
 
 $\mathcal{R}^*$ & Data Rate  & $1$ $Gbps$ to $10$ $Gbps$ \\ \hline
 $\mathcal{B}$ & Available Bandwidth & $10$ $GHz$\\ \hline
 
 $\mathcal{P}_o$ & Total Power & $0$ $dBm$ \\ \hline
 $\mathcal{P}_t$ & Transmit Power a& $\mathcal{P}_o/N$ \\ \hline
 $N$ & Number of THz-APs & $1,4,8,12,16$ \\ \hline
 
 $G (\delta)$ & Antenna Gain & $27$ $dBi$ \\ \hline
 $\delta$ & Antenna Beamwidth & $10\degree$\\ \hline

 $\mathcal{L}_T$ & Total Path Loss & -\\ \hline
 $f_c$ & Center Frequency  & $570$ $GHz$ \\ \hline
 $d_{ij}$ & Euclidian Distance & - \\ \hline
 
 $\rho$ & Relative Humidity  & $60\%$ \\ \hline
 $T$ & Room Temperature  & $25\degree C$ \\ \hline
 
 $N_f$ & Noise Power Spectral Density  & $-193.85$ $dB/Hz$ \\ \hline
 
 $\tau$ & Medium Absorption Coefficient  & - \\ \hline
 
 $t_{align}$ & Time for antenna alignment & $5$ $ms$ \\ \hline
 $L$ & Room Length & $10$ $m$ \\ \hline
 $h_r$ & Room Height & $3$ $m$ \\ \hline
 $h_c$ & Height Correction for Type C & - \\ \hline
 $M$ & Number of THz-UEs & $30$ \\ \hline
 $v$ & User Velocity & $1 \pm 0.5$ $m$ \\ \hline
 $h_u$ & Average User Height & $1.5$ $m$ \\ \hline
 $r_u$ & Average User Width & $0.2$ $m$ \\ \hline
 
 \hline
\end{tabular}
\label{Tab}
\end{table}

\subsection{Parameter Selection}

As shown in Equation \ref{RateEq}, the illumination intensity in a THz-Illumination region is dependent on multiple parameters. The choice of these parameters is critical for the efficiency of the AP placement types. In this example, the total path loss is dependent on four different factors, $f_c$, $d_{ij}$, $\rho$ and $T$. The last two factors, $\rho$ and $T$ are uncertain and varies with time and location. We assume it to be constant, as shown in Table \ref{Tab}. However, factors such as the operating frequency $f_c$ and the distance $d_{ij}$ can be adjusted based on the deployment strategy. By increasing the number of APs and decreasing the inter-AP distance, we can effectively reduce $d_{ij}$, since users are assigned based on the strongest link. The coverage and the density of the APs are also dependent on the choice of operating frequency. Higher THz frequency will result in higher path loss and reduced coverage area; thus, demanding more THz-AP redundancies.

Although lower THz frequencies will require lower redundancies, to reduce the THz-Darkness region in the lower frequency, we can also increase AP redundancies. To effectively increase the redundancy, we need to understand the coverage area of the THz-Illumination region. The illumination intensity is directly proportional to $\mathcal{B}$, which is immensely available in the THz. Still, to improve efficiency, we assume a channel bandwidth of $10$ $GHz$. Let $\mathcal{S}$ be the spectral efficiency for the THz-Illumination region of radius $r$. With simple manipulation on Equation \ref{RateEq} we have Equation \ref{ShanDist}.

\abovedisplayskip=0pt
\belowdisplayskip=4pt
\begin{eqnarray}
e^{\tau r}r^2=\frac{\mathcal{P}_oG_t (\delta)G_r (\delta)}{N_f(\frac{4\pi f_c}{c})^2(2^\mathcal{S}-1)}
\label{ShanDist}
\end{eqnarray}

Equation \ref{ShanDist} represents a popular mathematical form known as the Lambert's function \cite{Lambert}. Using  the Lambert's W-function given by the inverse of $f(\mathcal{W}) = \mathcal{W}e^\mathcal{W}$, we can compute the radius for the THz-Illumination region, shown in Equation \ref{OptiR}, where $K=\frac{\mathcal{P}_tG_t (\delta)G_r (\delta)}{N_f(\frac{4\pi f_c}{c})^2(2^\mathcal{S}-1)}$.

\abovedisplayskip=0pt
\belowdisplayskip=4pt
\begin{eqnarray}
 r = \ceil*{\frac{2\mathcal{W}(-\frac{\sqrt{K \tau^2}}{2})}{\tau}}
\label{OptiR}
\end{eqnarray}

\begin{figure}[t]
\centering
\includegraphics[width=3.5 in,height=2.5 in]{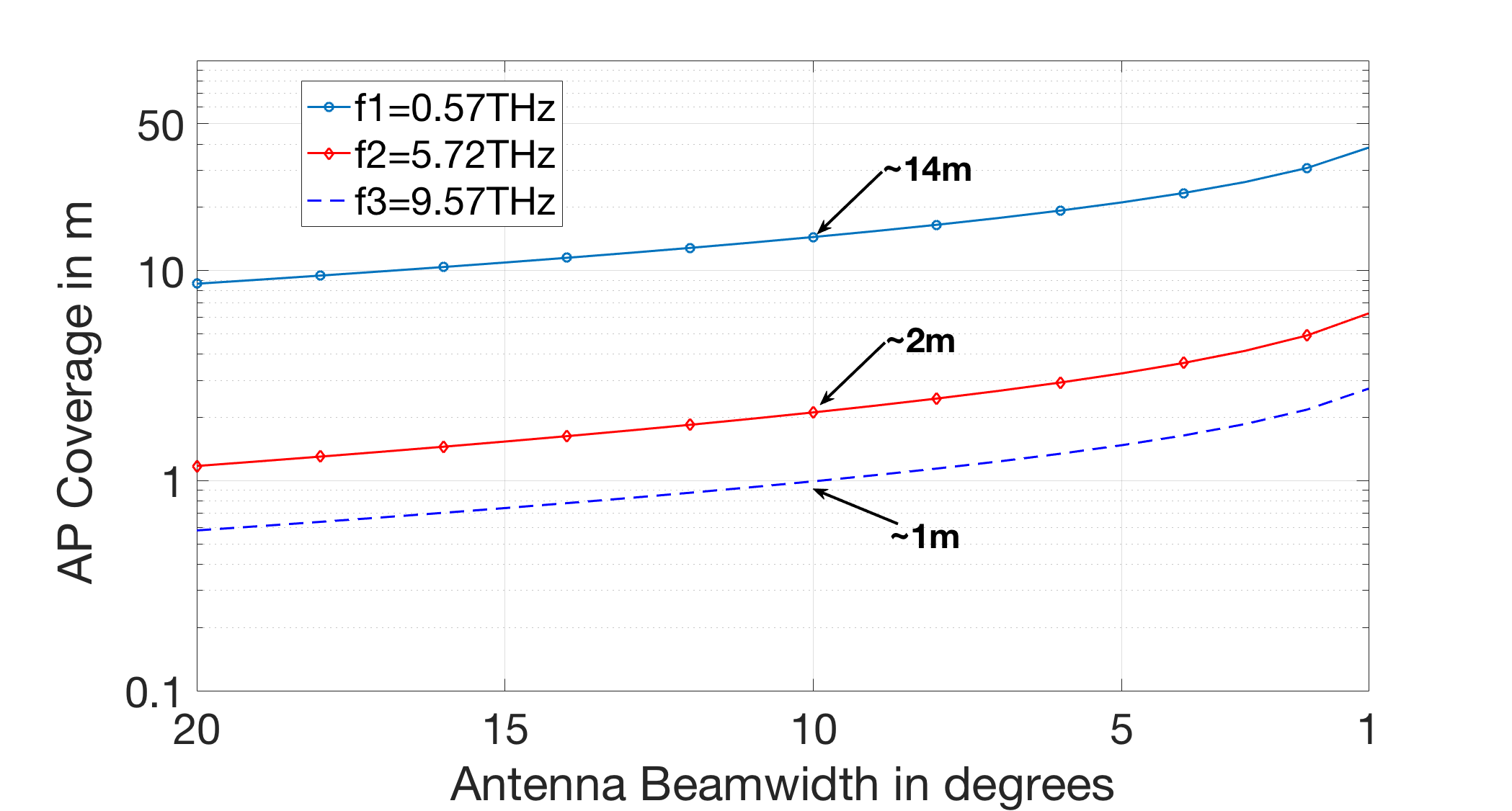}
\caption{THz-AP coverages at different frequencies to achieve minimum spectral efficiency of $0.1$ $Gbps/GHz$.}
\label{CovEval}
\end{figure}

Using the radius values shown in Equation\ref{OptiR}, we can observe the illumination range for a single THz-AP. Due to uncertainty in environmental parameters, the THz spectrum has multiple contiguous channel windows. Let us consider three different frequencies in the THz range, which have relatively larger contiguous bandwidth compared to other THz channels. Keeping the operating frequency fixed, we can improve the illumination range by using narrower antenna beamwidth, shown in Fig. \ref {CovEval}.

\setlength\belowcaptionskip{-0.2 in}
\setlength{\abovecaptionskip}{0 in}

\setlength\belowcaptionskip{-0.2 in}
\setlength{\abovecaptionskip}{0 in}
\begin{figure*}[t]
\centering
\includegraphics[width=6.5 in,height=4 in]{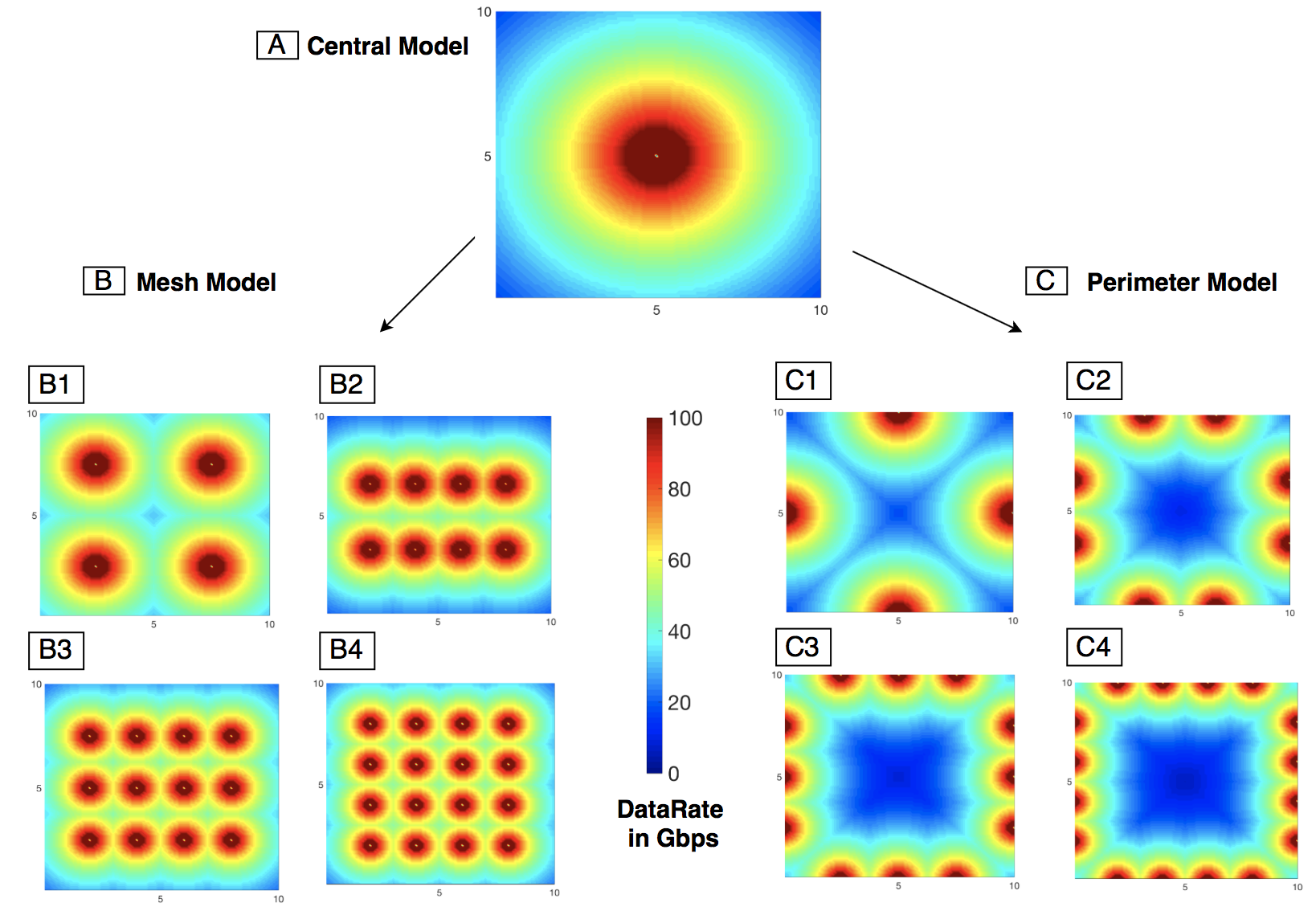}
\caption{Throughput heat map for varying number of THz-APs and placement types.}
\label{HM}
\end{figure*}

The downside to narrower beams is that the coverage area decreases, and the beams become highly sensitive to small-scale mobility \cite{SSMNano}\cite{OurCCNC} and blockage dimensions \cite{THzBlk}, resulting in multiple THz-Shadows. The antenna beamwidth either needs to be learned based on user mobility \cite{OurCCNC} or pre-adjusted based on analysis \cite{SSMNano}. For a room of $10m$-by-$10m$, the maximum possible separation between the AP-UE is $~14m$. For simplicity we fix the antenna beamwidth to $10\degree$ for frequency $f_1=570$ $GHz$. For simplicity we consider only frequency $f_1$ to evaluate SHINE placement methods, since for a room size of $10m$-by-$10m$ the frequencies $f_2$ and $f_3$ are unsuitable. These frequencies can be used for different use-cases, something in agreement with AP placement Type F. Detailed analysis for these frequencies is beyond the current scope of this paper.

\subsection{AP Placement Types} \label{ApModTy}

Although in Section \ref{SArch}, we proposed six different types of THz-AP placement strategies, there are two primary placement regions: (a) ceiling, and (b) perimeter. For simplicity, we compare three types of AP placement strategies: Type A (central), Type B (grid), and Type C (perimeter). Please note that Type D and E are derivatives of placement Type C, and Type F is a derivative of Type B. Moreover, using Type E and F placement with a room size of $10$ $m$-by-$10$ $m$ and frequency $f_1$ (relatively larger coverage area) will be a waste of resources.

To evaluate the impact of density on the THz-AP constellation, we consider multiple scenarios of THz-AP placement shown in Fig. \ref {HM}. Type A is a base case where there is only one THz-AP, which can access the whole of $\mathcal{P}_o$. We assume 4 sub-categories for Type B and C with $N= 4, 8,12,16$. The APs in Type B and C equally share the transmit power. Please note that in Fig. \ref {HM} all APs are placed at the same height of $h_r$

The data rate heat map, shown in Fig. \ref {HM}, demonstrates the extent of the THz-Illumination regions as defined in Equation \ref{OptiR}. Since THz-darkness depends on the choice of the bare minimum, let us assume the blue regions as the THz-darkness. Although Type B has a better THz-Illumination region, Type C placement has multiple advantages over Type B, which can be utilized to improve system efficiency. For example, Type B has a full $360\degree$ view, which results in a longer beam alignment time. On the other hand, Type C has only  a $180\degree$ view resulting in a $t_{align}/2$ beam alignment time. Moreover, Type B can also cause an unnecessary need for coverage and handoffs.

\setlength\belowcaptionskip{-0.2 in}
\setlength{\abovecaptionskip}{0 in}
\begin{figure}[t]
\centering
\includegraphics[width=3.5 in,height=2.5 in]{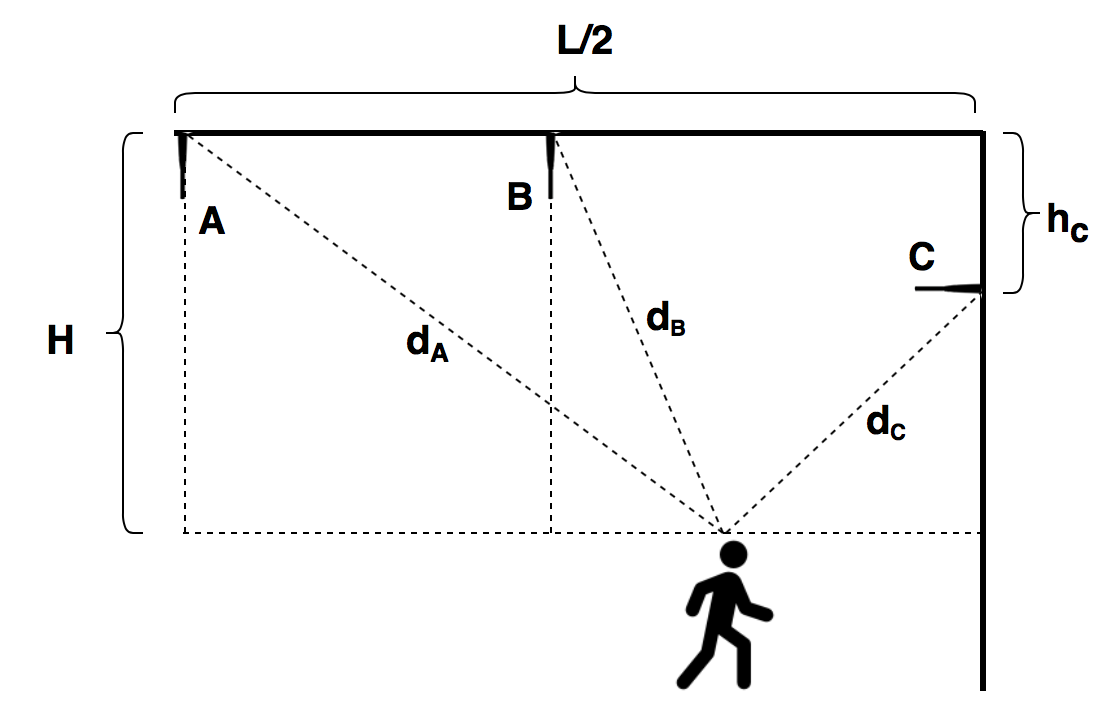}
\caption{User location relative to the AP placement types.}
\label{HcorSch}
\end{figure}

\setlength\belowcaptionskip{-0.2 in}
\setlength{\abovecaptionskip}{0.2 in}
\begin{figure*}[t]
\centering
\begin{subfigure}[]{2.2 in}
\includegraphics[width=2.3 in,height=2.5 in]{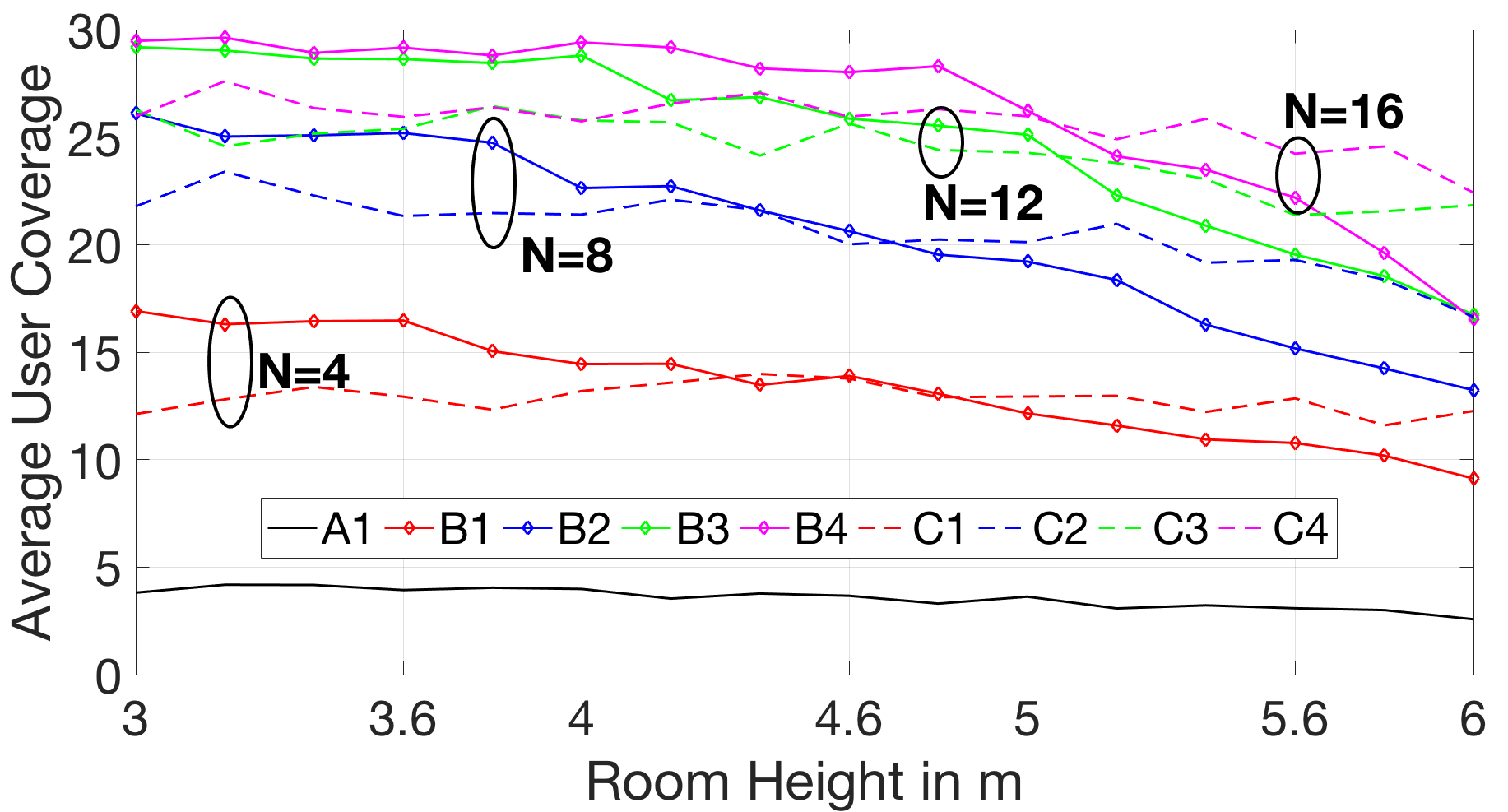}
\caption{User Coverage.}
\label{U1}
\end{subfigure}
~
\begin{subfigure}[]{2.2 in}
\includegraphics[width=2.3 in,height=2.5 in]{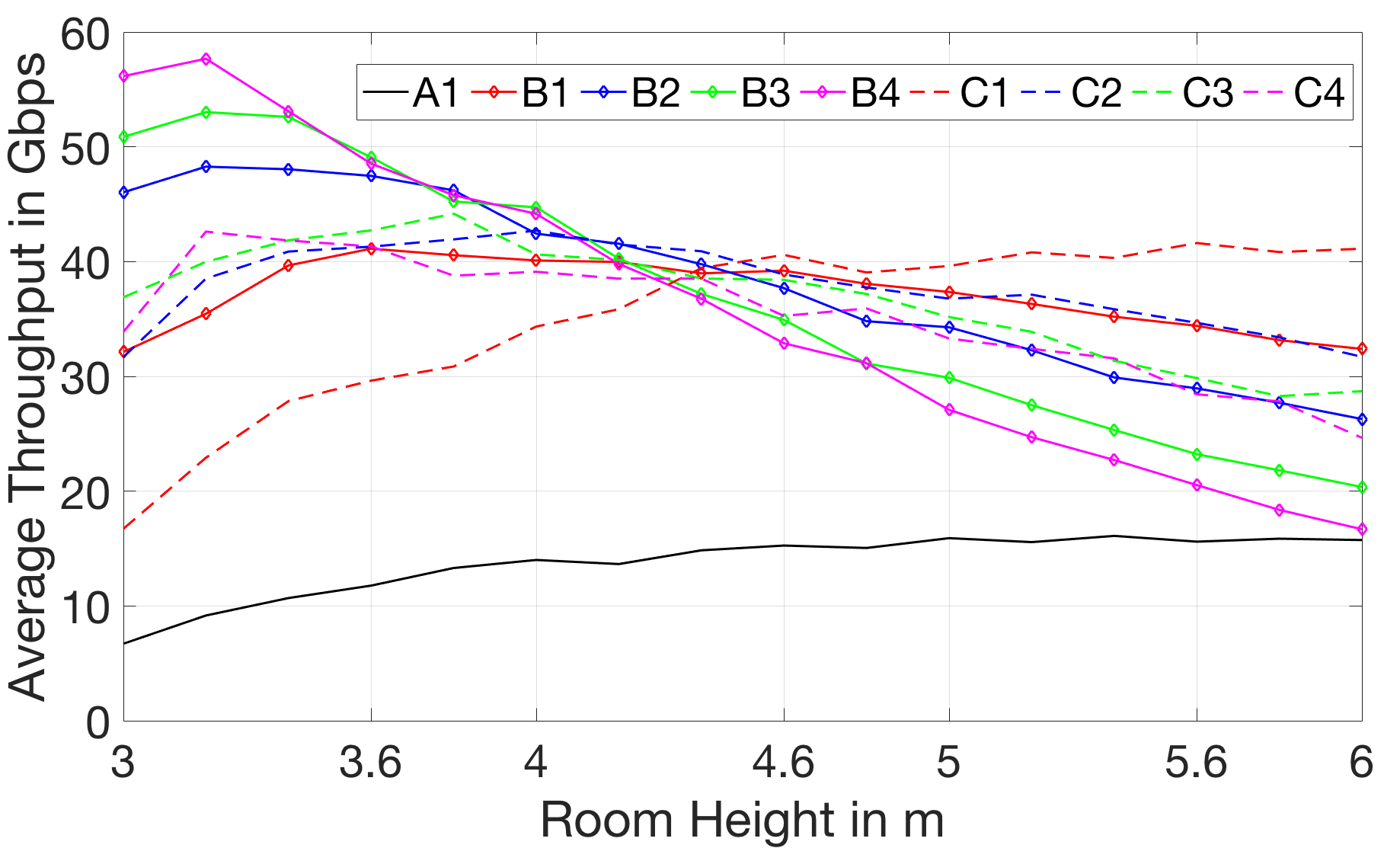}
\caption{Throughput.}
\label{T1}
\end{subfigure}
~
\begin{subfigure}[]{2.2 in}
\includegraphics[width=2.3 in,height=2.5 in]{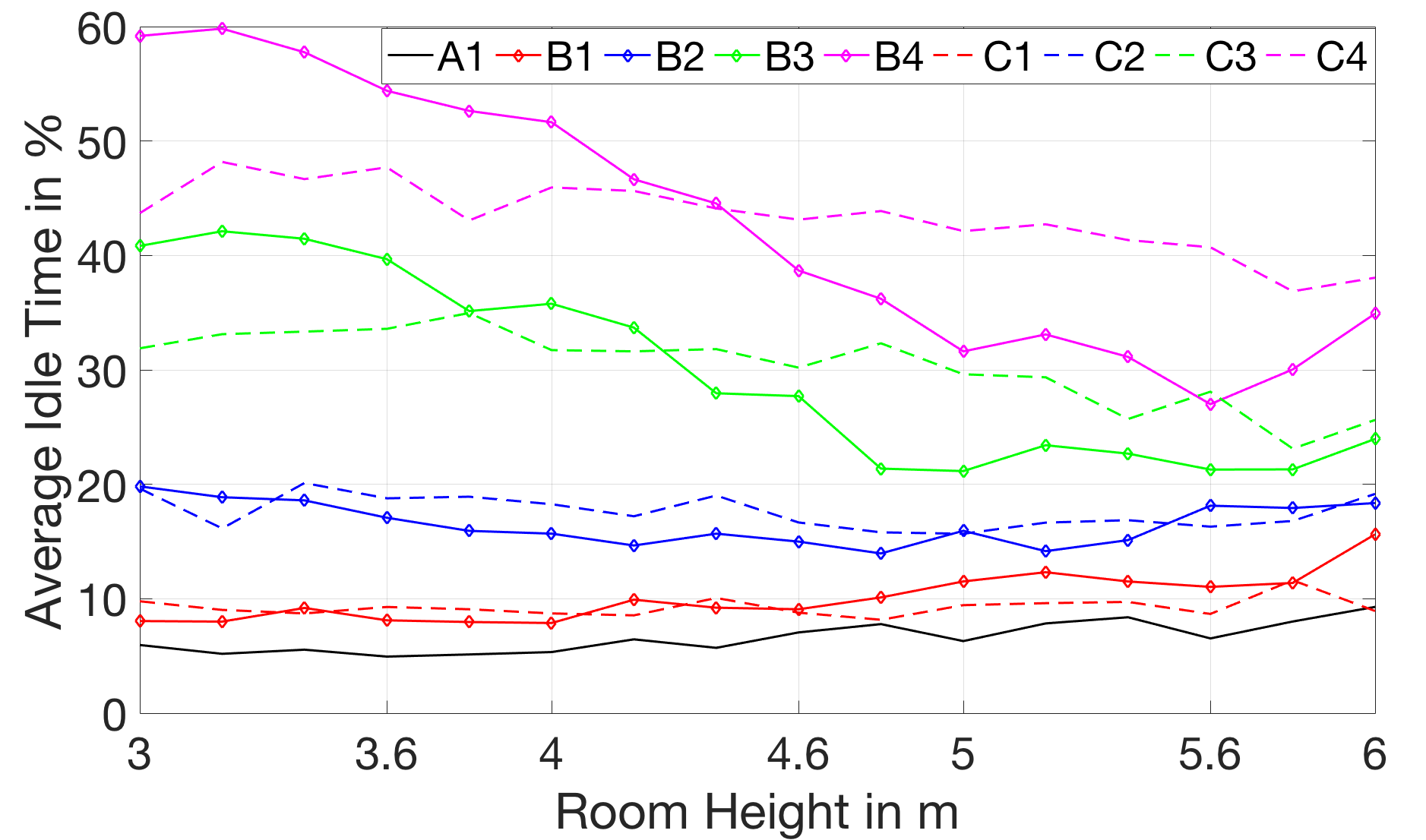}
\caption{AP Idle Time.}
\label{I1}
\end{subfigure}

\caption{Evaluation for varying number of THz-APs and AP placement types.} 
\label{Eval1}
\end{figure*}

\subsection{Height correction for Type C} \label{HRel}

The average distance between APs in Type B is $<L/4$ and a function of N, while for Type C its always $L$. Thus, at antenna height $h_r$ Type B is always a better choice compared to Type C, since Type C is farther away from the center. To access the advantages of Type C mentioned in Section \ref{ApModTy} we need to make Type B and Type C equivalent. One solution is to provide more transmit power to Type C APs, which is not possible due to the power constraint mentioned in Section \ref{SysModel}. Another solution is, to lower the height of Type C to compensate for the increased distance from the center.

The relative position of the user from the three types of APs are shown in Fig. \ref{HcorSch}. Let $H=h_r-h_u$ be the effective room height, which considers the distance between the THz-AP and the user device, which is assumed at the users\textquotesingle height. Let $h_c$ be the height correction factor for Type C APs that will make it equivalent to Type B APs in terms of Illumination. Using similarity of triangles we can show Equation \ref{SimiEq}.

\abovedisplayskip=-4pt
\belowdisplayskip=4pt
\begin{eqnarray}
\frac{d_B}{d_C}=\frac{H}{H-h_c}
\label{SimiEq}
\end{eqnarray}

Keeping other parameters constant for both Type B and C (i.e., $K_B=K_C$) and using Equation \ref{ShanDist}, we can show Equation \ref{CombEq}.

\abovedisplayskip=-4pt
\belowdisplayskip=4pt
\begin{eqnarray}
\frac{{d_B}^2}{{d_C}^2}=\frac{e^{\mathcal{K}d_C}}{e^{\mathcal{K}d_B}}
\label{CombEq}
\end{eqnarray}

Finally combining Equations \ref{SimiEq} \& \ref{CombEq}, we can solve for $h_c$ as shown in  Equation \ref{HcorEq}. Equation \ref{HcorEq} shows that the effective room height is a critical parameter for Types B and C. For the rest of the paper we assume that Type B and Type C along with height correction have same illumination intensity.

\abovedisplayskip=-4pt
\belowdisplayskip=4pt
\begin{eqnarray}
h_c=H(1-e^\frac{\tau (d_B-d_C)}{2})
\label{HcorEq}
\end{eqnarray}

 
\section{SHINE Evaluation} \label{SEval}

In this Section, we evaluate the SHINE model for multiple scenarios shown in Fig. \ref{HM} and parameters listed in Table \ref{Tab}. 

\subsection{Placement Evaluation}

We assume $M=30$ users, which move with a velocity of $v$ and follow a random-way point model \cite{OurCCNC}. Although a room can have a mixed AP deployment, for simplicity and better evaluation of each AP type, we assume a room is deployed with only one category of SHINE AP type, as shown in Figure \ref{HM}. Since the users are relatively mobile, we assume that the THz-APs have to realign before every transmission. Type A and B takes $t_{align}$ and Type C $t_{align}/2$ time for every realignment. The THz-APs handoff these mobile users based on the signal strength. In the evaluation shown in Fig \ref{Eval1}, we assume that there are no blockages, i.e., no THz-Shadow, and there is either a direct LOS with the users or NLOS through reflection.

Since it was shown in Section \ref{HRel} that the height of the AP is critical to the efficiency of the AP types proposed in SHINE, we evaluate our model for varying effective room height $H$. Type A is the baseline and has the lowest efficiency among all nine scenarios. As shown in Fig. \ref{Eval1}, the user coverage, throughput, and AP idle time decrease with increasing AP height. However, there are some cutoff points in height were Type C performs better than Type B. If we compare Type B and C in pairs based on the number of APs, i.e., 4 (red), 8 (blue), 12 (green), 16 (pink), we can observe a trend. For example, at effective room height of $4-5$ $m$ Type C is better than Type B for better user coverage, shown in Fig. \ref{U1}. Moreover, for higher room height, denser deployment of THz-APs are not that effective for the system. The system is better off with fewer number of THz-APs.

A similar trend can be observed for throughput, where the efficiency of the system flips around at $4.5$ $m$ height, i.e., Type C has a higher throughput than Type B APs, shown in Fig. \ref{T1}.  To measure energy utilization, we consider the idle time for all the APs, as shown in Fig. \ref{I1}. As the height increases, the Type A \& B APs are compelled to utilize more resources to cater to users, which is not the case for Type C.

\setlength\belowcaptionskip{-0.2 in}
\setlength{\abovecaptionskip}{0 in}
\begin{figure}[t]
\centering
\includegraphics[width=3 in,height=2 in]{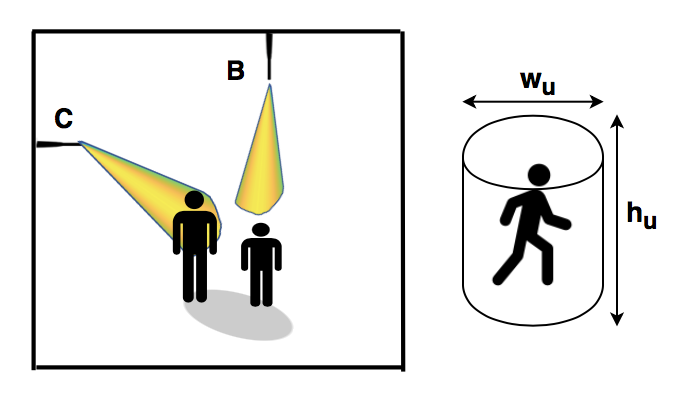}
\caption{Human Blockage modeling for THz-AP Types B \& C.}
\label{BlkMod}
\end{figure}

\setlength\belowcaptionskip{-0.2 in}
\setlength{\abovecaptionskip}{0.2 in}
\begin{figure*}[t]
\centering
\begin{subfigure}[]{2.2 in}
\includegraphics[width=2.3 in,height=2.5 in]{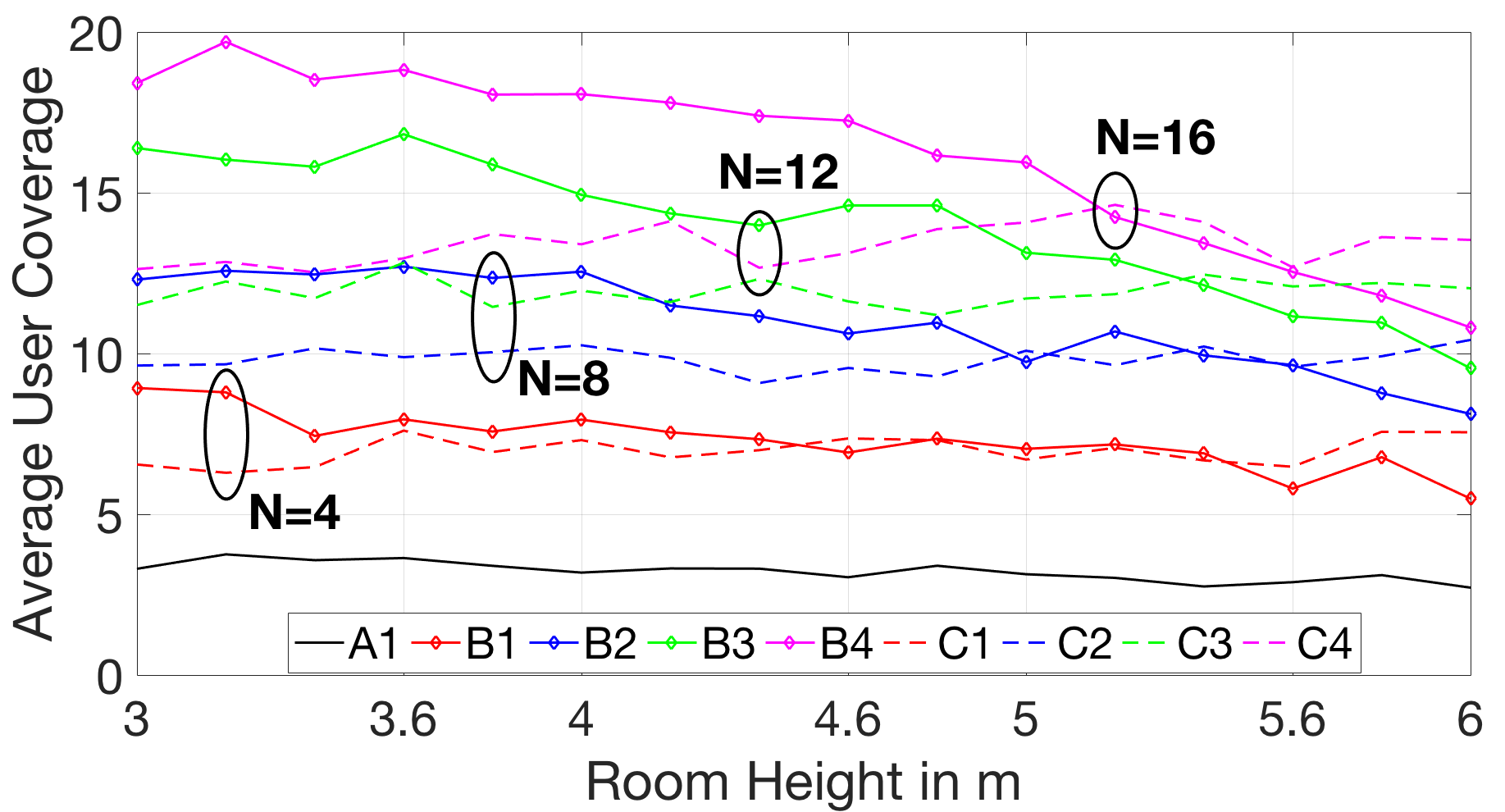}
\caption{User Coverage.}
\label{U2}
\end{subfigure}
~
\begin{subfigure}[]{2.2 in}
\includegraphics[width=2.3 in,height=2.5 in]{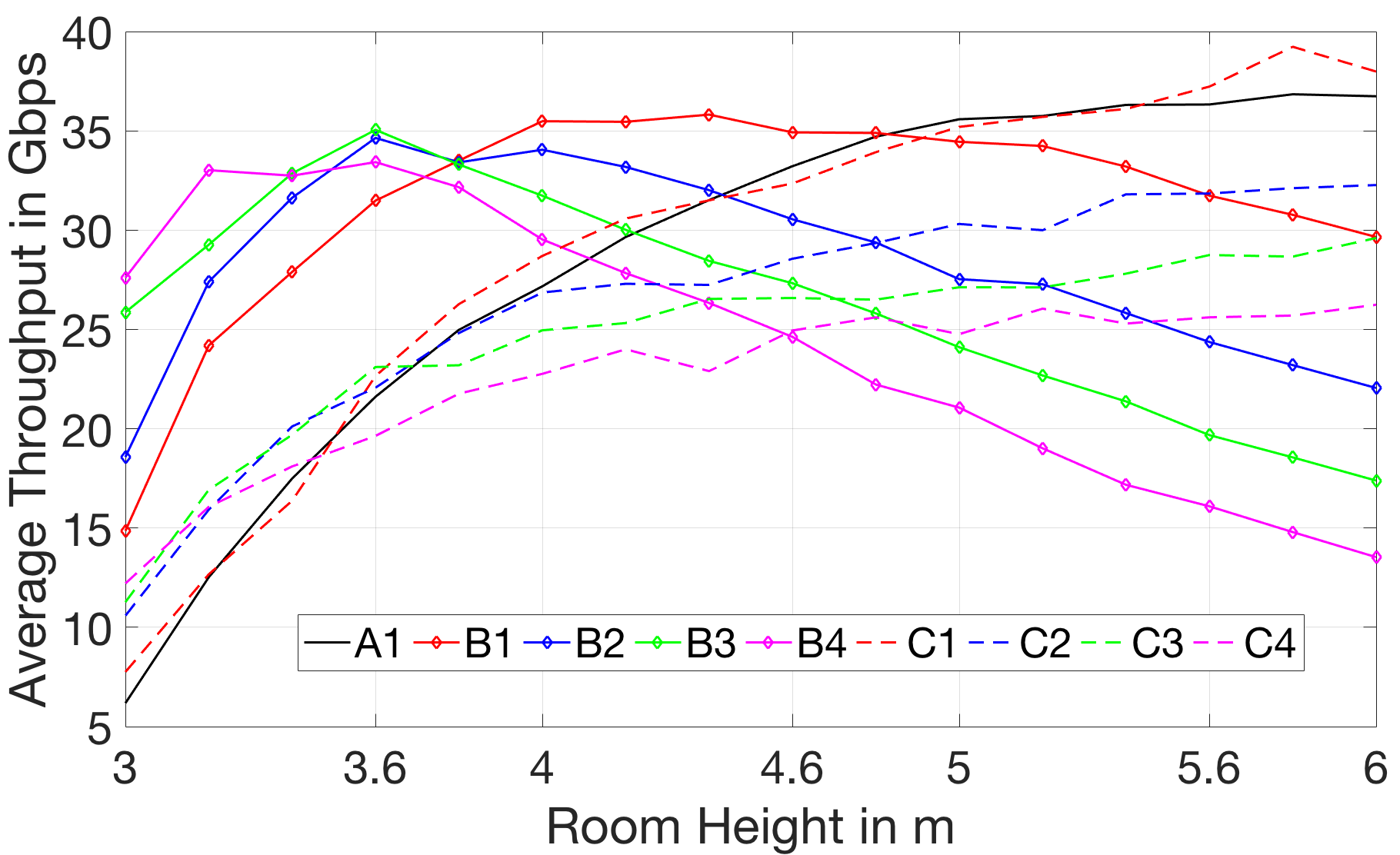}
\caption{Throughput.}
\label{T2}
\end{subfigure}
~
\begin{subfigure}[]{2.2 in}
\includegraphics[width=2.3 in,height=2.5 in]{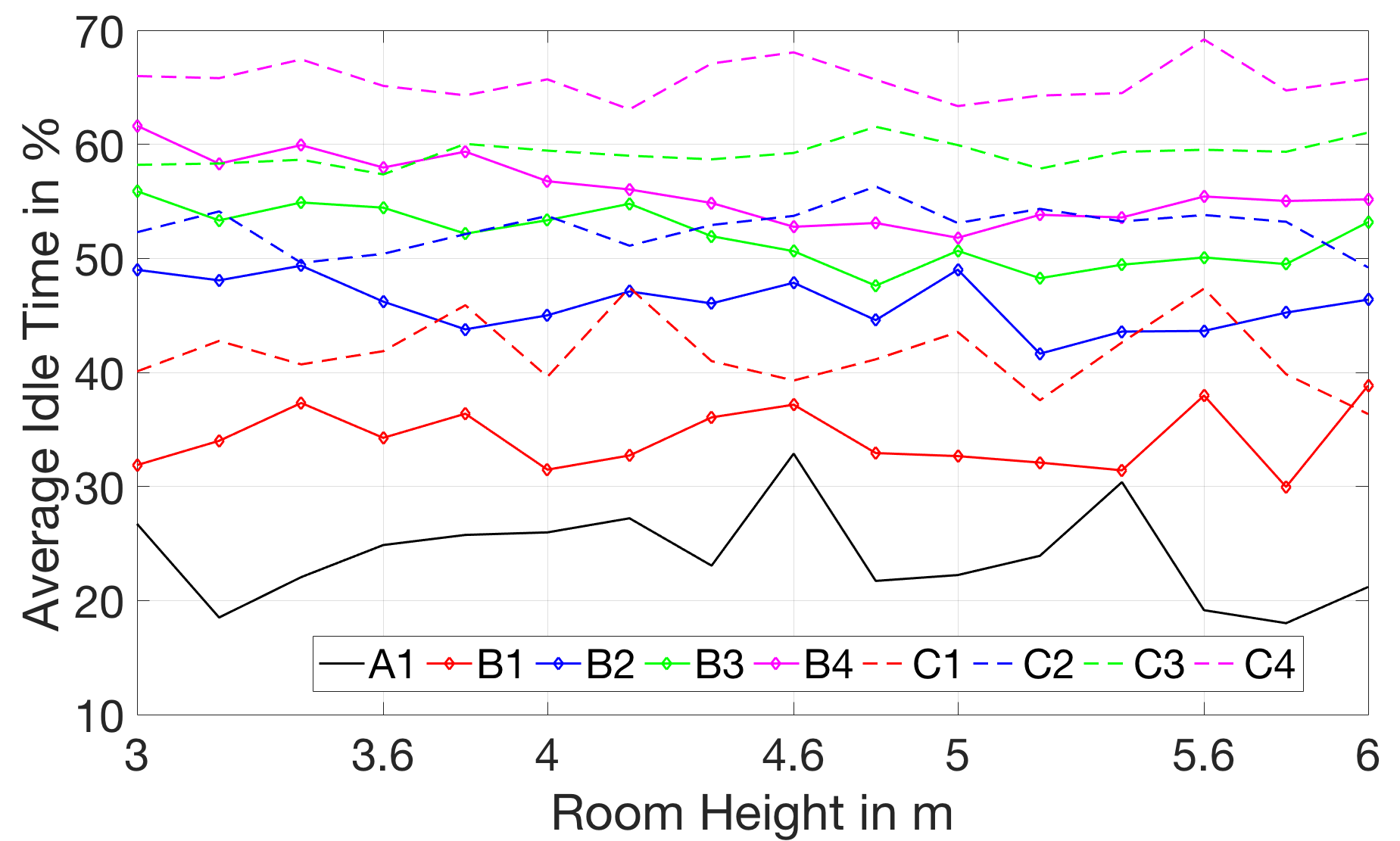}
\caption{AP Idle Time.}
\label{I2}
\end{subfigure}

\caption{Evaluation for varying number of THz-APs and AP placement types with mobile blockages.} \label{Eval2}
\end{figure*}

\subsection{Blockage Evaluation}

Till now, we have considered solutions to reduce THz-Darkness, but to reduce THz-Shadow, we need to track blockages and avoid mobility induced outages. With the proper deployment, static blockages can be avoided, but mobile blockages make the system unpredictable and unreliable. This compels the need for deploying multiple THz-APs, which might result in too much wasting of resources.

Furthermore, to compensate for the high path losses, it is necessary that we improve the antenna gains, i.e., make the antenna beams narrower. However, with narrower beams, the THz link becomes sensitive to blockage dimensions \cite{THzBlk}. Thus, every user will cast a THz-Shadow by default. For example, a user who is taller and broader can cast a big THz-Shadow on the user nearby, as shown in Fig. \ref{BlkMod}. However, the strategic placement of THz-APs will help mitigate these blockage related challenges. For example, for the scenario shown in Fig. \ref{BlkMod} Type B has a better THz-Illumination footprint compared to Type C. Moreover, Type C has faster beam alignment, which might give an edge over Type B in some scenarios. To incorporate these mobile blockages in our evaluation we consider user height $h_u$ and width $r_u$ as shown in Table \ref{Tab}. Although the efficiency of the system decrease with introduction of blockages, as shown in Fig. \ref{Eval2}, we can still observe cutoff points where Type C placement is better than Type B. Please note the variation in Fig. \ref{I2} is due to the random movement pattern of the mobile blockages, i.e., the UEs.

Although in most cases, Type B outperforms Type C due to its uniform distribution, Type C can be used for specific uses-cases to improve coverage and avoid human blockages.  For example, irrespective of the number of APs, in rooms with higher elevation, Type C placement performs better than Type B placement. In this paper, we assumed that the height of the room is uniform throughout. However, in the future, we can consider a mixed distribution of THz-APs with a portion of the room using Type B and the other portion using Type C THz-APs. For example, rooms having indoor stairs will have mixed SHINE architecture, where the region near the stairs will have Type C APs, and the rest of the room can have Type B APs.


\section{Conclusion} \label{Con}

In this paper, through SHINE, we presented multiple strategies for deploying THz-APs efficiently in a room and proposed models through which these APs can blend in with the current indoor lighting systems. SHINE introduced that with proper AP constellation, density placement, and limited energy, we can significantly improve the THz-Illumination region or system efficiency. Instead of having a single THz-AP at the center of the room, it will be beneficial to divide the single AP resource into multiple smaller THz-APs, as shown through Type B and C.  However, we also show that there is limit to this densification of THz-APs depending on the type of the AP, room dimension, and user mobility. A higher density of APs does not necessarily mean better coverage. User mobility and its relative location with the APs can impact the THz-Illumination and THz-Shadow regions. In the future, we will evaluate the other types of THz-AP placement introduced in this paper and also consider a mixed AP placement for specific use cases.

\end{document}